\documentclass{sigchi-ext}
\usepackage[T1]{fontenc}
\usepackage{textcomp}
\usepackage[scaled=.92]{helvet} 
\usepackage{graphicx} 
\usepackage{balance}  
\usepackage{booktabs} 
\usepackage{ccicons}  
\usepackage{ragged2e} 

\def\plaintitle{Responsible AI and Its Stakeholders} 

\def\emptyauthor{}
\def\plainkeywords{Artificial Intelligence; Responsibility; Responsible AI; Blameworthiness; Accountability; Liability}

\title{Responsible AI and Its Stakeholders}

\numberofauthors{2}

\author{%
  \alignauthor{%
    \textbf{Gabriel Lima}\\
    \affaddr{School of Computing} \\
    \affaddr{KAIST} \\
    \email{gabriel.lima@kaist.ac.kr} }\alignauthor{%
    \textbf{Meeyoung Cha}\\
     \affaddr{Data Science Group}\\
    \affaddr{Institute for Basic Science (IBS)}\\
    \email{mcha@ibs.re.kr}
    } 
}

\definecolor{linkColor}{RGB}{6,125,233}
\hypersetup{%
  pdftitle={\plaintitle},
  pdfauthor={\emptyauthor},
  pdfkeywords={\plainkeywords},
  bookmarksnumbered,
  pdfstartview={FitH},
  colorlinks,
  citecolor=black,
  filecolor=black,
  linkcolor=black,
  urlcolor=linkColor,
  breaklinks=true,
}

\begin{document}

\setcopyright{rightsretained}
\copyrightinfo{\acmcopyright}

\maketitle

\RaggedRight{} 

\begin{abstract}
    Responsible Artificial Intelligence (AI) proposes a framework that holds all stakeholders involved in the development of AI to be responsible for their systems. It, however, fails to accommodate the possibility of holding AI responsible per se, which could close some legal and moral gaps concerning the deployment of autonomous and self-learning systems. We discuss three notions of responsibility (i.e., blameworthiness, accountability, and liability) for all stakeholders, including AI, and suggest the roles of jurisdiction and the general public in this matter. 

\end{abstract}




\section{Introduction}

\textit{Responsible Artificial Intelligence} (AI) is an approach that aims to consider the ethical, moral, and social consequences during the development and deployment of AI systems~\cite{dignum2019responsible}. Given the broad impact of AI on the future society, discussing the responsibility of different stakeholders involved in its implementation is essential. The current discussion, however, disregards the possibility of holding the AI itself responsible for its actions.

Scholars have discussed various ethical and legal gaps that might arise with the deployment of AI. 
For instance, the \textit{responsibility}~\cite{matthias2004responsibility} and \textit{accountability gaps}~\cite{koops2010bridging} are created by the unpredictability of self-learning AI systems and their distance to the persons that may employ them, impeding the assignment of moral responsibility and liability to users and manufacturers. On the other hand, the \textit{retribution gap} raises the question of who will be proper subjects of retributive blame as neither AI systems, users, and developers might be so~\cite{danaher2016robots}.

This work intends to discuss the possibility of holding the AI itself responsible for its actions, alongside other stakeholders like users and manufacturers. We tackle backward-looking meanings of responsibility proposed in van de Poel et al.~\cite[p.12-49]{van2015moral}, which focus on the evaluation of past actions and assignments of blame, accountability, and liability. 

We focus on
the attribution of responsibility for wrongful actions or omissions with negative consequences, rather than beneficial outcomes. For instance, the first pedestrian fatality caused by an autonomous car resulted in Uber (i.e., the owner and co-manufacturer) paying the price for the accident. While the safety operator of the car was the main actor to blame for the accident~\cite{crashuber}, Uber took the whole responsibility-as-liability, namely the duty to remedy the consequences of the accident. 
Could the autonomous car or its AI have been held responsible for the accident, together with the operator and the co-manufacturer?

\section{Responsibility as Blameworthiness}

Responsibility-as-blameworthiness proposes that agent \textit{i} should be held responsible if it is appropriate to attribute blame to \textit{i} for a specific action or omission. Blame assignment is restricted in
its primary objective as individual victims or patients choose to assign blame to \textit{i} as a form of retribution. While not extensive, the following conditions have been argued to be necessary for responsibility-as-blameworthiness~\cite{van2018varieties}: 1) moral agency, 2) causality, 3) knowledge, 4) freedom, and 5) wrongdoing. 

All stakeholders in the development of AI are defined to be \textit{moral agents}. Corporations, while not moral agents by themselves, are collective groups of moral agents. Without delving into the discussion of whether there exists free will, all stakeholders are considered to be \textit{free} unless they are mentally-impaired or under-aged. The main concern of attributing blame to developers revolves around the \textit{causality} and \textit{knowledge} conditions. Scholars have discussed whether the actions of a highly autonomous and self-learning artificial agent could break the chain of causation~\cite{chopra2011legal}. Innovation, especially self-learning AI and robots, also involves a great deal of uncertainty and unpredictability, conflicting with the knowledge condition~\cite{ turner2018robot}. Given this paper's focus on wrongful actions, all existing stakeholders are subject to wrongdoing.

AI, on the other hand, is not a moral agent. The discussion around the possibility of embedding moral values into AI and treating it as an entity with its own moral compass has scholars on both extremes of the spectrum of support~\cite{wallach2008moral} and opposition~\cite{bryson2010robots}. 
While the causality condition is easy to tackle as the AI is indeed the entity causally responsible for the wrongdoing, all other conditions cannot be satisfied by the fact that an AI is not a moral agent and does not understand its actions. 

Nonetheless, blame is attributed by actors who might choose to assign it regardless of the fulfillment of the conditions discussed above. Previous work has shown that humans are retributivists and look for someone to blame~\cite{cushman2008crime}. Blame assignment was found to be a two-step process initiated by the causal inspection of the wrongful action, followed by an analysis of the mental state of the agent. Even though AI might not be aware of its actions, humans might choose to attribute responsibility-as-blameworthiness to AI due to its causal connection to the negative consequences.

\section{Responsibility as Accountability}

An agent \textit{i} is considered responsible-as-accountable for a specific action had \textit{i} been assigned the role to bring about or to prevent it. In contrast to the blame assignment, holding an agent accountable only requires 1) the agent's capacity to act responsibly and 2) a causal connection between \textit{i} and the action~\cite{van2015moral}. 

The concept of Responsible AI proposes that all stakeholders in the development and deployment of AI should act responsibly to the best of their ability. As developers, manufacturers, and users can always be traced back to moral agents, they are arguably capable of acting responsibly. However, due to the self-learning and unpredictability of self-learning AI, the capacity of responsible action of these stakeholders might be somewhat hindered. 
As discussed in the previous section, attributing ``causality to either the physical person or company that is behind the (electronic) agent'' might also become difficult as AI becomes more autonomous and distributed~\cite{koops2010bridging}.

An AI is developed to perform (or prevent) specific actions from happening, which could, at first thought, imply that these systems could be easily held accountable by definition. 
Additionally, the causal connection between the wrongdoing and the AI is easy to determine. However, AI cannot act responsibly if it does not understand what acting responsibly means as it does not comprehend the moral consequences of its actions, as pointed out when discussing responsibility-as-blameworthiness.




\section{Responsibility as Liability}

Responsibility-as-liability has juridical and legal decisions as to its premise. Here, we focus on the attribution of liability regardless of moral agency, as legal systems often do through strict liability assignment, for instance. The duty of liability to agent \textit{i} implies that \textit{i} should remedy or compensate certain parties for its action or omission. 

As in the case of Uber and its autonomous vehicle that caused the death of a pedestrian, corporations are currently often held liable for wrongdoings of its AI systems. Exemplifying this trend, Volvo has promised to take full responsibility, namely liability, for its self-driving cars~\cite{volvo}. While holding existing legal persons liable for the actions of AI might promote safe systems in the short term, it could hurt innovation and adoption in the long run~\cite{turner2018robot}.

Holding AI liable for its actions would require a legal reframing of the legal status of AI, i.e., the adoption of electronic legal personhood. The European Parliament~\cite{eurecommendation} has previously considered such possibility, initiating controversial debate among scholars~\cite{van2018we, bryson2017and}. The extension of such legal status to AI would also require these systems to hold their own assets~\cite{turner2018robot} or insurance premiums~\cite{vladeck2014machines} so they can compensate those harmed and remediate the consequences. Previous work has also found a public desire to attribute liability to AI even though people are aware that these entities do not satisfy such preconditions for punishment~\cite{lima2020explaining}.





\section{Concluding Remarks}

The concept of \textit{Responsible AI} aims to promote the responsible development and deployment of AI systems through the assignment of responsibility to all stakeholders involved in the process. It neglects, however, the possibility of holding these systems themselves responsible for their actions. While AI is considered not a moral nor a responsible agent, it could be held responsible by jurisdiction or the general public.

These newly developed and ever innovating AI systems challenge various notions of responsibility, creating legal and moral gaps in society~\cite{matthias2004responsibility, koops2010bridging, danaher2016robots}. These gaps cannot be solved solely by holding users and manufacturers responsible for the actions of AI due to the difficulty of satisfying the knowledge and causality requirements of responsibility assignment. The latter, for instance, can be easily satisfied by an AI had it been the entity that caused the wrongful action. Additionally, the general public might find AI blameworthy for its actions as a result of human retributivism. Therefore, society in the future might consider holding autonomous AI systems responsible alongside other stakeholders.

Holding AI responsible for its actions is not a comprehensive solution to the issues discussed above. While attributing responsibility to AI might solve some of these gaps, it also raises various questions, such as which legal status should be granted and how an AI would compensate those harmed. Nevertheless, the concept of Responsible AI stresses a framework that holds mainly developers and manufacturers blameworthy, accountable, and liable for the actions of AI, challenging the very concept its name might suggest: holding AI responsible per se.




\balance{} 

\scriptsize{
\bibliographystyle{SIGCHI-Reference-Format}
\bibliography{sample}


\begin{thebibliography}{00}


\ifx \showCODEN    \undefined \def \showCODEN     #1{\unskip}     \fi
\ifx \showDOI      \undefined \def \showDOI       #1{{\tt DOI:}\penalty0{#1}\ }
  \fi
\ifx \showISBNx    \undefined \def \showISBNx     #1{\unskip}     \fi
\ifx \showISBNxiii \undefined \def \showISBNxiii  #1{\unskip}     \fi
\ifx \showISSN     \undefined \def \showISSN      #1{\unskip}     \fi
\ifx \showLCCN     \undefined \def \showLCCN      #1{\unskip}     \fi
\ifx \shownote     \undefined \def \shownote      #1{#1}          \fi
\ifx \showarticletitle \undefined \def \showarticletitle #1{#1}   \fi
\ifx \showURL      \undefined \def \showURL       #1{#1}          \fi

\bibitem{eurecommendation}
 2017.
\newblock {\em European Parliament report with recommendations to the
  Commission on Civil Law Rules on Robotics (2015/2103(INL))}.
\newblock


\bibitem{volvo}
{Allianz Partners} (Ed.). 2018.
\newblock {\em Self-driving cars: Volvo to take full responsibility for all
  accidents}.
\newblock
\newblock
\shownote{Available at \url{https://tinyurl.com/wwzv2rw}. Date accessed
  29/01/2020.}


\bibitem{bryson2010robots}
{Joanna~J Bryson}. 2010.
\newblock \showarticletitle{Robots should be slaves}.
\newblock {\em Close Engagements with Artificial Companions: Key social,
  psychological, ethical and design issues\/} (2010), 63--74.
\newblock


\bibitem{bryson2017and}
{Joanna~J Bryson}, {Mihailis~E Diamantis}, {and} {Thomas~D Grant}. 2017.
\newblock \showarticletitle{Of, for, and by the people: the legal lacuna of
  synthetic persons}.
\newblock {\em Artificial Intelligence and Law\/} {25}, 3 (2017), 273--291.
\newblock


\bibitem{chopra2011legal}
{Samir Chopra} {and} {Laurence~F White}. 2011.
\newblock {\em A legal theory for autonomous artificial agents}.
\newblock University of Michigan Press.
\newblock


\bibitem{cushman2008crime}
{Fiery Cushman}. 2008.
\newblock \showarticletitle{Crime and punishment: Distinguishing the roles of
  causal and intentional analyses in moral judgment}.
\newblock {\em Cognition\/} {108}, 2 (2008), 353--380.
\newblock


\bibitem{danaher2016robots}
{John Danaher}. 2016.
\newblock \showarticletitle{Robots, law and the retribution gap}.
\newblock {\em Ethics and Information Technology\/} {18}, 4 (2016), 299--309.
\newblock


\bibitem{dignum2019responsible}
{Virginia Dignum}. 2019.
\newblock {\em Responsible Artificial Intelligence: How to Develop and Use AI
  in a Responsible Way}.
\newblock Springer International Publishing.
\newblock


\bibitem{koops2010bridging}
{Bert-Jaap Koops}, {Mireille Hildebrandt}, {and} {David-Olivier
  Jaquet-Chiffelle}. 2010.
\newblock \showarticletitle{Bridging the accountability gap: Rights for new
  entities in the information society}.
\newblock {\em Minn. JL Sci. \& Tech.\/}  {11} (2010), 497.
\newblock


\bibitem{crashuber}
{Dave Lee}. 2019.
\newblock {\em Uber self-driving crash 'mostly caused by human error'}.
\newblock
\newblock
\shownote{Available at \url{https://tinyurl.com/r8klk59}. Date accessed
  29/01/2020.}


\bibitem{lima2020explaining}
{Gabriel Lima}, {Meeyoung Cha}, {Chihyung Jeon}, {and} {Kyungsin Park}. 2020.
\newblock \showarticletitle{Explaining the Punishment Gap of AI and Robots}.
\newblock {\em arXiv preprint arXiv:2003.06507\/} (2020).
\newblock


\bibitem{matthias2004responsibility}
{Andreas Matthias}. 2004.
\newblock \showarticletitle{The responsibility gap: Ascribing responsibility
  for the actions of learning automata}.
\newblock {\em Ethics and information technology\/} {6}, 3 (2004), 175--183.
\newblock


\bibitem{turner2018robot}
{Jacob Turner}. 2018.
\newblock {\em Robot Rules: Regulating Artificial Intelligence}.
\newblock Springer.
\newblock


\bibitem{van2015moral}
{Ibo Van~de Poel}, {Lamb{\`e}r~MM Royakkers}, {Sjoerd~D Zwart}, {and} {Tiago
  De~Lima}. 2015.
\newblock {\em Moral responsibility and the problem of many hands}.
\newblock Routledge New York.
\newblock


\bibitem{van2018varieties}
{Ibo van~de Poel} {and} {Martin Sand}. 2018.
\newblock \showarticletitle{Varieties of responsibility: Two problems of
  responsible innovation}.
\newblock {\em Synthese\/} (2018), 1--19.
\newblock


\bibitem{van2018we}
{Robert van den~Hoven van Genderen}. 2018.
\newblock \showarticletitle{Do We Need New Legal Personhood in the Age of
  Robots and AI?}
\newblock In {\em Robotics, AI and the Future of Law}. Springer, 15--55.
\newblock


\bibitem{vladeck2014machines}
{David~C Vladeck}. 2014.
\newblock \showarticletitle{Machines without principals: liability rules and
  artificial intelligence}.
\newblock {\em Wash. L. Rev.\/}  {89} (2014), 117.
\newblock


\bibitem{wallach2008moral}
{Wendell Wallach} {and} {Colin Allen}. 2008.
\newblock {\em Moral machines: Teaching robots right from wrong}.
\newblock Oxford University Press.
\newblock


\end{thebibliography}
}

\end{document}